# Reduction of friction by normal oscillations.
# II. In-plane system dynamics

Xinyu Mao[1,2], Valentin L. Popov[1,3,4*], Jasminka Starcevic[1,4], Mikhail Popov[1,3,4]

[1]Technische Universität Berlin, Berlin 10623, Germany
[2]Tsinghua University, Beijing 100084, China
[3]Tomsk Polytechnic University, Tomsk 634050, Russia
[4]Tomsk State University, Tomsk 634050, Russia

* Corresponding author: Valentin L. Popov
E-Mail address: v.popov@tu-berlin.de
Telephone number: +49 (030) 314-21480

**Abstract:** The influence of out-of-plane oscillations on friction is a well-known phenomenon that has been studied extensively with various experimental methods, e.g. pin-on-disk tribometers. However, existing theoretical models have yet achieved only qualitative correspondence with experiment. Here we argue that this may be due to the system dynamics (mass and tangential stiffness) of the pin or other system components being neglected. This paper builds on the results of a previous study (Popov M. et al. *Friction*, 2016, submitted) by taking the stiffness and resulting dynamics of the system into account. The main governing parameters determining macroscopic friction, including a dimensionless oscillation amplitude, a dimensionless sliding velocity and the relation between three characteristic frequencies (that of externally excited oscillation and two natural oscillation frequencies associated with the contact stiffness and the system stiffness) are identified. In the limiting cases of a very soft system and a very stiff system, our results reproduce the results of previous studies. In between these two limiting cases there is also a resonant case, which is studied here for the first time. The resonant case is notable in that it lacks a critical sliding velocity, above which oscillations no longer reduce friction. Results obtained for the resonant case are qualitatively supported by experiments.

## 1 Introduction

Vibrations can be applied to reduce and control friction, which is widely used in many industrial branches, such as metal forming, wire drawing and drilling [1], [2]. One of the earliest studies of friction reduction due to oscillations was carried out by Godfrey in 1967 [3]. He conducted experiments, in which a rider slid along a steel plate and was vibrated in the direction perpendicular to the plane. Afterwards numerous studies were carried out, which can be roughly classified by whether the static or sliding friction is considered and by the direction of the oscillations, see e.g. [4], [5], [6]. The three possible directions of oscillation are: (1) in the sliding direction; (2) perpendicular to the sliding direction in the contact plane; (3) perpendicular to the contact plane (out-of-plane oscillations). Arbitrary combinations of these three modes are also possible, some of which can produce directed motion even in the absence of a directed mean force, thus producing a frictional drive. In this regard, active control of friction through oscillations is closely related to oscil-lation-based frictional drives [7], [8]. However, in the present paper we consider only sliding friction under the influence of out-of-plane oscillations.

Friction under the action of out-of-plane oscillations has been studied experimentally in the past in a number of works [9], [10], [11], [12]. The first theoretical description was proposed in [13],[14], where the movement of a rigid body under constant tangential force and oscillating normal force was considered. Unfortunately this model achieved only qualitative correspondence with experimental results. In [7], it was shown that the macroscopic behavior of a frictional contact is strongly influenced by the contact stiffness. Related studies of the dependence of friction on tangential oscillations [15], [16] and a study of frictional drives [17] came to the same conclusion. In a recent experimental study [18] the contact stiffness was also confirmed as one of the main parameters governing the response of a tribological contact to high frequency oscillations.

Based on these indications of the importance of contact stiffness, M. Popov et al. [19] carried out a theoretical study of friction under the action of out-of-plane oscillations with explicit account of finite contact stiffness. In this paper it was assumed that the stiffness of the system is much larger than that of the contact, which allowed avoiding consideration of system dynamics. In real systems, depending on their particular mechanical design, the stiffness of the system may be comparable with the contact stiffness, thus bringing the whole system dynamics into play. In the present paper, we extend the previous study [19] by considering the complete dynamics of a system with a tribological contact.

## 2  Simplified model of the experimental set-up

The model studied in the present paper is motivated by experimental studies of active control of friction by out-of-plane oscillations in a pin-on-disk tribometer (e.g. [7], [14], [15], [16]). The design of the pin is shown in Fig. 1a. Assuming that the vertical stiffness of the set-up is much larger than the normal contact stiffness, the vertical macroscopic motion of the pin can be considered to be displacement controlled. The tangential stiffness of the pin assembly is much smaller than its vertical stiffness, so that it is no longer guaranteed that the tangential stiffness of the pin is larger than the tangential contact stiffness. Therefore, the tangential stiffness of the pin is explicitly taken into account in our model. Assuming that the transversal dynamics of the pin is controlled by only one bending normal mode of the pin, we arrive at the simplified model of the system, which is sketched in Fig. 1b: a one-degree-of-freedom model taking into account the normal and tangential contact stiffness, the inertia of the pin and its tangential stiffness. Modal analysis of the pin could be used for estimation of a more accurate modal mass, but we do not do this here, as our aim is to present a high-level analysis without considering particular geometrical realizations. We will show that the frequency of free oscillations of the pin, $\omega_0 = \sqrt{k_x/m}$, is the most important system parameter; by analyzing particular experiments, it has to be adjusted to the ground frequency of the free oscillations of the pin. Naturally, our model abstracts away many (possibly important) aspects of real frictional systems, in particular the differential contact stiffness of curved bodies (we model the contact as a single spring with constant stiffness). However, in the first part of this series [19] we found that the detailed contact mechanics had surprisingly little influence on the results, relative to a one-spring model. In particular, abstracting the exact geometry of the contact does not change the relevant dimensionless variables. Due to this, and in view of the already large number of system parameters we restrict ourselves to the simple model described above.

The model, as shown in Fig. 1b, consists of a rigid body with mass $m$ that is connected to an external actuator, which imposes the body's $z$-coordinate. The body is pulled by a spring with a tangential stiffness $k_x$ and interacts with the substrate through a "contact spring" that has the nor-mal stiffness $k_{z,c}$ and tangential stiffness $k_{x,c}$. The vertical movement of the mass is determined explicitly by the external oscillation:

$$u_z = u_{z,0} + \Delta u_z \cos \omega t \tag{1}$$

where $u_{z,0}$ is a constant initial indentation, $\Delta u_z$ is the amplitude of normal oscillations, and $\omega$ is the angular frequency of the oscillation. The attached "system spring" is pulled tangentially with a constant velocity $v_0$. The motion of the body in the $x$-direction under the influence of the attached springs is described by Newton's Second Law for the tangential displacement $u_x$. The tangential displacement of the immediate contact point is denoted with $u_{x,c}$. For simplicity, we assume Coulomb's law of friction with a constant coefficient of friction $\mu_0$ in the immediate contact point between the substrate and the contact spring. Although this may be an unrealistic assumption in general, the aim of this study is to understand how changes in the *macroscopic* coefficient of friction can arise from pure system dynamics even with *constant microscopic* friction. Experimental results might be best approximated by a combined theory, including system dynamics, contact mechanics and changes of the local coefficient of friction, but this is left for later studies.

Note that the amplitude of oscillation can be either smaller than the mean indentation (non-jumping), in which case the body is always in contact with the substrate, or larger (jumping case), where contact with the substrate is intermittent. Initially we will focus on the permanent contact case. Jumping will be introduced later in the paper.

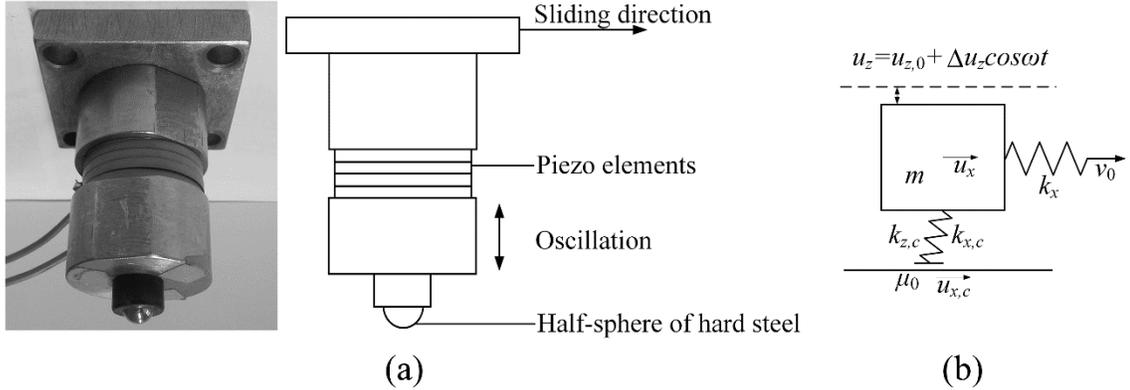

(a)          (b)

**Fig. 1**. (a) Photograph and diagram of the pin assembly of a pin-on-disk tribometer used in [7], [16] and in the experimental part of the present study; (b) A simplified model of the pin in contact with the disk, which is studied in the present paper.

## 3 Qualitative analysis

All previous studies of the influence of normal oscillations on friction, including the first part of the present work [19], have shown qualitatively the same dependence of the macroscopic coefficient of friction (COF) on velocity: At zero velocity the friction force is at its static value, which is determined solely by the minimum value of normal force during one oscillation cycle:

$$\mu(v_0 = 0) = \begin{cases} \mu_0 \left(1 - \Delta u_z / u_{z,0}\right), & \text{for } \Delta u_z < u_{z,0} \quad \text{(non-jumping case)} \\ 0, & \text{for } \Delta u_z > u_{z,0} \quad \text{(jumping case)} \end{cases}. \tag{2}$$

At higher velocities the COF increases until it reaches $\mu_0$ at some critical velocity $v_0^*$ ("point of insensitivity"), and remains constant thereafter. The static COF and the critical velocity $v_0^*$ are the two main reference points of the velocity dependence of the COF. While the first reference point is universal (2), the second one is determined by the dynamics of the tribological system.

We begin our analysis with the case of small oscillation amplitudes, $\Delta u_z < u_{z,0}$, so that the lower point of the indenter remains in contact with the substrate at all times. In this case, the normal component of the contact force $F_N$ is non-vanishing and is determined by the product of the current indentation depth (1) with the normal contact stiffness $k_{z,c}$:

$$F_N = k_{z,c}\left(u_{z,0} + \Delta u_z \cos \omega t\right). \tag{3}$$

At sufficiently large pulling velocities $v_0$, the contact point will be sliding all the time (without stick) in one direction (except for the resonant case that will be described later). Under these conditions, the average tangential force is equal to the average normal force times $\mu_0$, and the macroscopically observed COF, which we define as the ratio of the mean values of tangential and normal forces over one period of oscillation, will be constant and equal to $\mu_0$. When the above conditions are satisfied the tangential force of the contact spring is in equilibrium with the friction force (normal force times $\mu_0$), since the contact stiffness is not associated with any mass:

$$k_{x,c}(u_x - u_{x,c}) = \mu_0 k_{z,c}\left(u_{z,0} + \Delta u_z \cos \omega t\right). \tag{4}$$

The equation of motion of the body $m$ along the $x$-axis is

$$m\ddot{u}_x = k_x(v_0 t - u_x) - k_{x,c}(u_x - u_{x,c}). \tag{5}$$

Taking (4) into account, the equation of motion reads

$$m\ddot{u}_x + k_x u_x = k_x v_0 t - \mu_0 k_{z,c} \Delta u_z \cos \omega t - \mu_0 k_{z,c} u_{z,0}. \tag{6}$$

The particular solution of this equation is

$$u_x = v_0 t - \mu_0 \frac{k_{z,c}}{k_x} u_{z,0} + \frac{\mu_0 k_{z,c} \Delta u_z}{m\omega^2 - k_x} \cos \omega t. \tag{7}$$

Differentiating this solution with respect to time gives

$$\dot{u}_x = v_0 - \frac{\mu_0 k_{z,c} \Delta u_z}{m\omega^2 - k_x} \omega \sin \omega t. \tag{8}$$

Differentiating the equilibrium condition (4),

$$k_{x,c}(\dot{u}_x - \dot{u}_{x,c}) = -\mu_0 k_{z,c} \Delta u_z \omega \sin \omega t, \tag{9}$$

substituting (8) into (9) and resolving the resulting equation with respect to $\dot{u}_{x,c}$, we obtain the following expression for the sliding velocity of the immediate contact point (lower point of the contact spring)

$$\dot{u}_{x,c} = v_0 - \frac{k_{z,c}}{k_{x,c}} \frac{k_{x,c} + k_x - m\omega^2}{(m\omega^2 - k_x)} \mu_0 \omega \Delta u_z \sin \omega t . \tag{10}$$

Due to our previous assumption of continuous sliding this velocity must remain positive at all times. This is the case if

$$v_0 > v_0^*, \tag{11}$$

where

$$v_0^* = \mu_0 \omega \Delta u_z \frac{k_{z,c}}{k_{x,c}} \frac{|k_{x,c} + k_x - m\omega^2|}{|k_x - m\omega^2|}. \tag{12}$$

This relatively simple equation is one of the central results of the present paper and it is instructive to discuss it in some detail. First, let us consider limiting cases that have already been studied in the literature:

I. In the case of a very soft system ($k_x \ll m\omega^2$) with very large contact stiffness ($k_{x,c} \gg m\omega^2$) we effectively have a rigid body under the action of constant tangential force. In this case, which was considered in [14] (see esp. Fig. 20 and discussion) the critical velocity reduces to

$$v_{0,soft}^* = \mu_0 \omega \frac{\Delta u_z k_{z,c}}{|m\omega^2|} = \mu_0 \frac{\Delta F_N}{m\omega}. \tag{13}$$

II. The limiting case of a very stiff system ($k_{x,c} \ll |k_x - m\omega^2|$) was considered in the first part of the present work, [19]. In this case (12) simplifies to

$$v_{0,stiff}^* = \mu_0 \omega \Delta u_z \frac{k_{z,c}}{k_{x,c}}. \tag{14}$$

There are two other limiting cases which involve resonances and have not yet been considered in the literature:

III. If $k_{x,c} + k_x - m\omega^2 \approx 0$, the critical velocity is very small: $v_0^* \approx 0$. The body is in permanent sliding state even at very low velocities and the COF is constant and equal to $\mu_0$ at all sliding velocities.

IV. If $k_x - m\omega^2 \approx 0$, the critical velocity is infinitely large and the system never achieves the state of continuous sliding. It will be shown that in this case the macroscopic coefficient of friction reaches a plateau at large velocities, with a value lower than $\mu_0$. This case is of a special interest and it will be considered below in detail and was also studied experimentally.

Let us now consider the movement of the body in the general case, when the contact point slides during some part of the oscillation cycle and sticks at other times. The movement of the slider is still governed by the Eq. (5), however, Eq. (4), which describes the tangential force in the contact spring, is only valid during the sliding part of the period, while during the sticking phase the following is true for the immediate contact point:

$$\dot{u}_{x,c} = 0, \quad k_{x,c}(u_x - u_{x,c}) < \mu_0 k_{z,c} \left( u_{z,0} + \Delta u_z \cos \omega t \right). \tag{15}$$

To study the dynamics of the system in detail, the equation of motion (5) was integrated numerically with account of (4) and (15). The nontrivial behavior that can result when both stick and slip occur is illustrated in Fig. 2, which presents the time dependencies of the normal and tangential force (the former multiplied with $\mu_0$). During the sliding phase (e.g. before $t_1$ and between $t_2$ and $t_3$), these two quantities are equal, while during the sticking phase the tangential force (green line) is less than the normal force times $\mu_0$ (blue line). The beginning of stick ($t_1$) is determined by the condition that the velocity of the immediate contact point (lower end of the contact spring) becomes zero, while the end of the sticking phase ($t_2$) is determined by the condition that the tangential force becomes equal to the normal force times $\mu_0$.

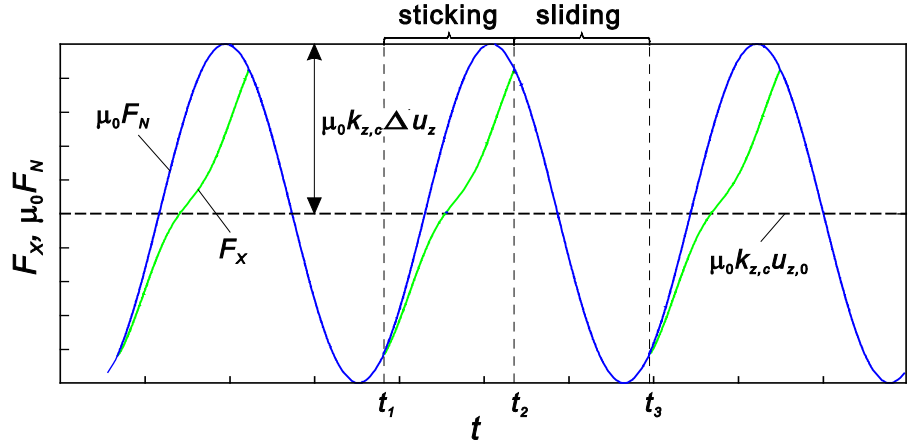

**Fig. 2.** An example of the dynamics of normal and tangential contact forces showing the phases of slip, where the tangential force coincides with the normal force multiplied with the COF, and the sticking phase where the tangential force (green line) is smaller than the normal force multiplied with the COF (blue line).

## 4 Dimensionless formulation of the problem

Introducing the dimensionless variables

$$\bar{v} = \frac{v_0}{v_0^*}, \tag{16}$$

$$\tau = \omega t, \tag{17}$$

$$\xi = u_x \frac{\omega}{v_0^*}, \tag{18}$$

$$\xi_c = u_{x,c} \frac{\omega}{v_0^*}, \tag{19}$$

where $v_0^*$ is defined by (12). With two additional dimensionless parameters

$$\alpha = \frac{k_x}{m\omega^2}, \quad \beta = \frac{k_x + k_{x,c}}{m\omega^2}, \tag{20}$$

we can rewrite the Eqs. (4), (5) and (15) in the following form:

$$(\xi - \xi_c) = \left(\frac{u_{z,0}}{\Delta u_z} + \cos\tau\right)\frac{\alpha - 1}{\beta - 1} \tag{21}$$

$$\xi'' = \alpha(\bar{v}\tau - \xi) - (\beta - \alpha)(\xi - \xi_c), \tag{22}$$

$$\xi'_c = 0, \quad \xi - \xi_c < \left(\frac{u_{z,0}}{\Delta u_z} + \cos\tau\right)\frac{\alpha - 1}{\beta - 1} \tag{23}$$

where $\xi' = d\xi/d\tau$, $\xi'' = d^2\xi/d\tau^2$.

One can see that the behavior of the above system is unambiguously determined by the following set of variables:

$$\bar{v}, \; \Delta u_z / u_{z,0}, \; \alpha, \; \text{and} \; \beta. \tag{24}$$

After solving the Eqs. (21)-(23), one can go back to the initial dimensional variables and calculate the average normal force $\langle F_N \rangle$ and the average tangential force $\langle F_x \rangle = \langle k_x (v_0 t - u_x) \rangle$. The macroscopic coefficient of friction is then defined as

$$\mu_{\text{macro}} = \frac{\langle F_x \rangle}{\langle F_N \rangle}. \tag{25}$$

It is easy to see that with the given dimensionless variables (24) the macroscopic coefficient of friction will be proportional to $\mu_0$. Thus, it is more convenient to define the reduced coefficient of friction, $\mu_{\text{macro}} / \mu_0$, which is a function solely of the variables (24). In the following, we will explore the dependence of the reduced COF on the dimensionless velocity (16) on the parameter plane $(\alpha, \beta)$.

## 5 Numerical results and analysis

We begin with a general classification of the numerical results (Fig. 3). According to the definition (20), $\beta$ is always larger than $\alpha$, therefore we only consider the upper half of the parameter space above the line $\alpha = \beta$. In the figure, it is easy to identify the previously described special cases: The limiting case of a very soft system with a stiff contact (case I, according to the above classification) corresponds to small values of $\alpha \ll 1$ and large values of $\beta \gg 1$, and is thus to be found in the upper left corner of the diagrams. The limiting case of very stiff system with low contact stiffness (case II), corresponds to $\alpha \approx \beta$ and is found along the diagonal of the diagram. The resonant case III corresponds to the line $\beta = 1$ and the resonant case IV to $\alpha = 1$.

Since case II occupies the diagonal of the diagram, there are infinitely many possible transitions from II to I. We will consider two such transitions: between A and B in Fig. 3, which passes over the resonant case III and from C to B, which passes over the resonant case IV.

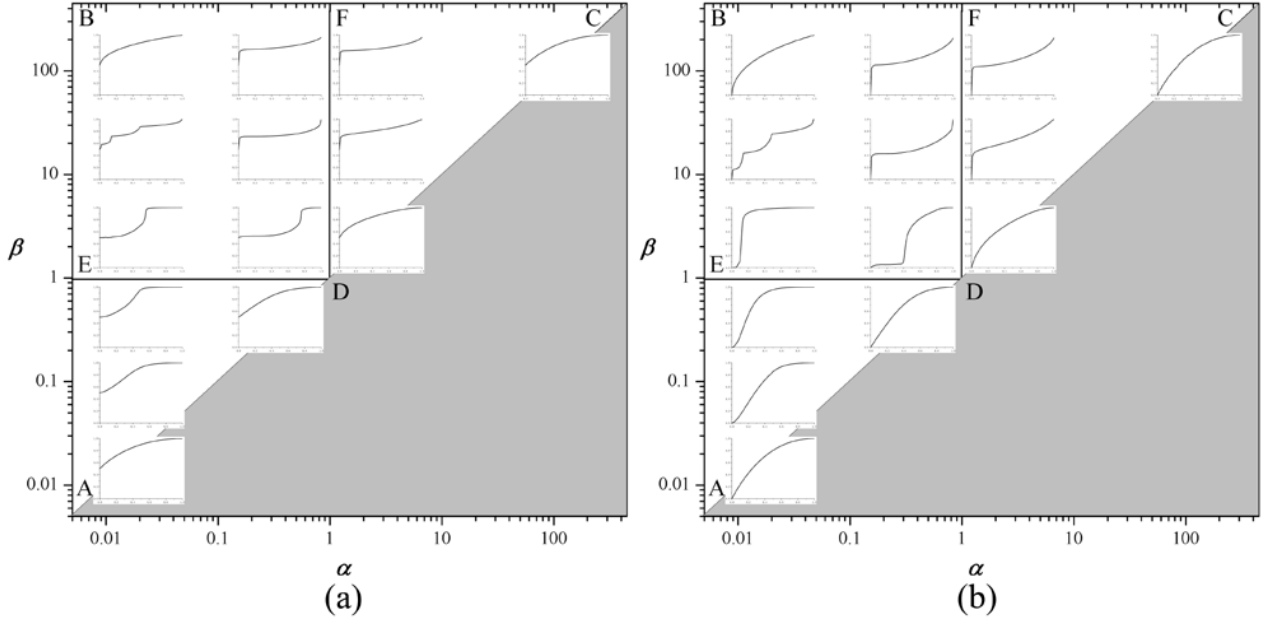

**Fig. 3.** Typical dependencies of $\mu_{\text{macro}}/\mu_0$ on $\bar{v}$ for the oscillation amplitude $\Delta u_z/u_{z,0} = 0.5$ (a) and for the maximal non-jumping oscillation amplitude $\Delta u_z/u_{z,0} = 1.0$ (b) arranged in a matrix of the dimensionless parameters $\alpha$ and $\beta$. The individual curves start at the static COF value at $\bar{v} = 0$, which only depends on the oscillation amplitude and is equal to $\mu_0/2$ in (a) and zero in the diagrams in (b). With increasing velocity, the reduced COF monotonically increases and reaches the value "1" at the velocity $\bar{v} = 1$. In between, however, the velocity-dependence of the COF is determined by the particular system dynamics.

## Limiting cases of soft (case I) and stiff (case II) system and transition over resonant case III

Let us consider the transition from the stiff to the soft system over the resonant case III in more detail. We start with the separate consideration of the limiting cases of very stiff and very soft system. The diagram in Fig. 4a shows results of numerical simulation for the parameter set $(\alpha,\beta) = (0.01, 0.02)$, which corresponds to the limiting case I according to the classification of Section 3. This case was considered in detail in the publication [19]. In Fig. 4, results of numerical simulations are compared with the semi-empirical equation

$$\frac{\mu_{\text{macro}}}{\mu_0} \approx 1 - \frac{\Delta u_z}{u_{z,0}}\left[\frac{3}{4}(\bar{v}-1)^2 + \frac{1}{4}(\bar{v}-1)^4\right] \tag{26}$$

derived in [19] with $\bar{v}$ given by (16) and $v^*$ by (14). The numerical data practically ideally coincide with the analytical result (26).

The right-hand-side diagram Fig. 4b presents a comparison for the opposite case of very soft system. Again, numerical data are compared with the analytical ones

$$\frac{\mu_{\text{macro}}}{\mu_0} = \left(1 - \frac{\Delta F_N}{F_{N,0}}\right) + \frac{\Delta F_N}{F_{N,0}}\left[\sqrt{\frac{4\pi}{9}}\,\bar{v} + \left(1 - \sqrt{\frac{4\pi}{9}}\right)\bar{v}^{1,2}\right] \tag{27}$$

obtained in [14], with $v^*$ given by (13). In this case too we see a very good agreement. However, numerical data have a noticeable fine structure which the limiting case curves do not have (a sort of small amplitude oscillations).

With these two limiting cases, we ensure the connection to the former studies of the problem and at the same time pose a more general problem of investigation of dependencies of the coefficient of friction on velocity between these two poles.

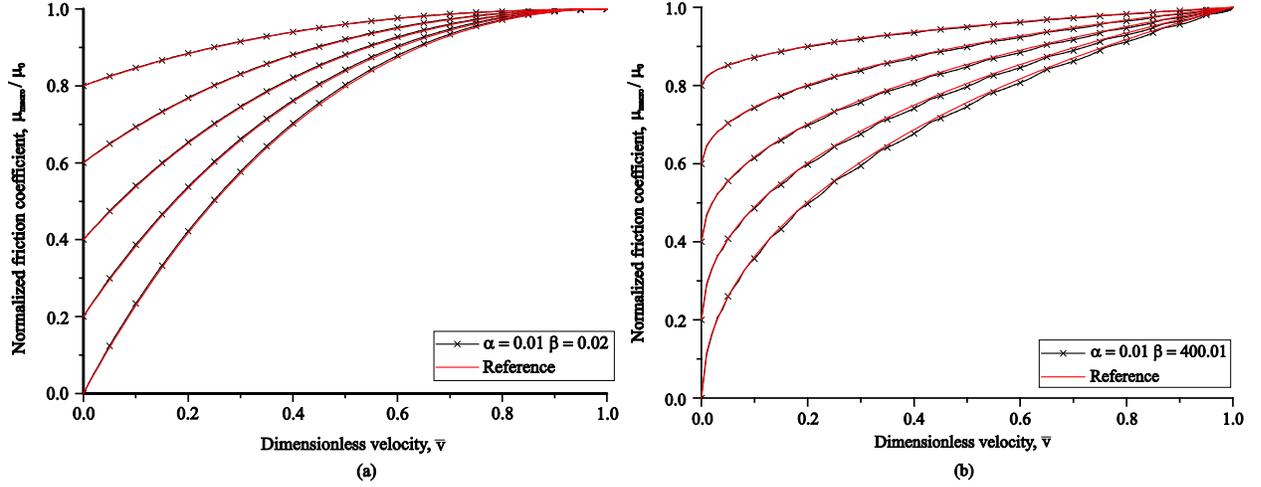

**Fig. 4.** The dependence of $\mu_{macro} / \mu_0$ on $\overline{v}$ for $(\alpha, \beta) = (0.01, 0.02)$ and $(\alpha, \beta) = (0.01, 400.01)$ and for the oscillation amplitudes $\Delta u_z / u_{z,0} = 0.2, 0.4, 0.6, 0.8, 1.0$ (from top to bottom). The crosses and black lines represent results of numerical simulation and the red lines the analytical results (26) and (27).

As character of the transformation of the law of friction is very similar for various oscillation amplitude, in the following we illustrate this transformation only for the case of the critical amplitude $\Delta u_z / u_{z,0} = 1$. The transition from the lower left corner of the diagram Fig. 3 to the upper left corner means that the value of the parameter $\alpha$ remains small, while parameter $\beta$ is changing from very small to very large values. The corresponding transformation of the dependencies of the reduced coefficient of friction on the dimensionless velocity is shown in Fig. 5 a for the values of $\beta$ in the lower left quarter of the parameter space and in Fig. 5b for the values of $\beta$ in the upper left quarter of the diagram. In the lower quarter, the changes of the form are relatively slow until parameter $\beta$ becomes very near to the value of "1". In the vicinity of this "resonant value" the upper point of the diagram starts to expand to the left forming a plateau (as is well seen in Fig.5a for $\beta = 0.91$). In the exact resonant case, the whole "dependence" consist only of this single plateau, that means that the coefficient of friction is constant and equal to $\mu_0$. Much more dramatic changes occur after passing the resonant value $\beta = 1$. The resonant plateau then sharply decreases while at the same time appears the next plateau. This process repeats many times producing an Oscillating curve which "upper envelope" tends toward the limiting solution for the soft system as already have been shown in Fig. 4b.

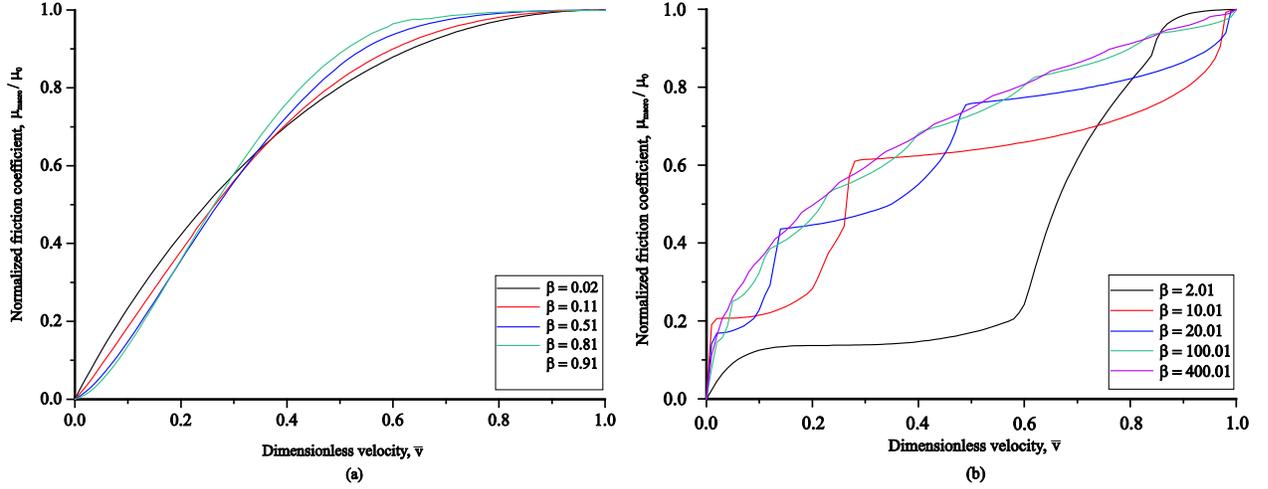

**Fig. 5** The dependence of $\mu_{macro}/\mu_0$ on $\bar{v}$ for $\Delta u_z/u_{z,0}=1$, $\alpha=0.01$ and a series of $\beta$: (a) lower left quarter of the diagram, (b) upper left quarter of the diagram.

*Resonant case IV*

We now turn our attention to the resonant case IV, where the frequency of oscillation is equal to the natural frequency of the slider $\omega=\sqrt{k_x/m}$ and the system's behavior is qualitatively different from previously considered cases. For convenience we consider an equivalent system, where the right end of the system spring $k_x$ in Fig. 1 is fixed and the substrate is instead moving with velocity $v_0$. Note also, that the velocity dependence of the COF cannot be displayed as a function of $\bar{v}$ in the resonant case, because $v^*$ tends to infinity and $\bar{v}$ becomes zero. We therefore return to the ordinary dimensional variables in this section. Since we are here concerned with very large values of $\beta$, we can consider the contact stiffness to be infinitely large for the purposes of this analysis. We chose the direction of the x-axis as the direction of movement of the substrate. The equation of motion then reads

$$m\ddot{x}+k_x x=\mu_0 F_N \mathrm{sign}(v_0-\dot{x}). \tag{27}$$

In our model the normal force oscillates according to

$$F_N=F_{N,0}+\Delta F\cdot\cos\omega t. \tag{28}$$

Thus, the complete equation of motion is

$$m\ddot{x}+k_x x=\mu_0\left(F_{N,0}+\Delta F\cdot\cos\omega t\right)\mathrm{sign}(v_0-\dot{x}). \tag{29}$$

For an approximate analysis, let us assume that the body begins with a small-amplitude oscillation

$$\dot{x}=\Delta v\cos\omega t \tag{30}$$

Then the amplitude will be increasing over time until $\Delta v$ becomes larger than $v_0$. Indeed, multiplying (29) with (30) and noting that the left-hand side of the resulting equation is the time derivative of the energy of the system, we arrive at the energy equation

$$\frac{\mathrm{d}}{\mathrm{d}t}\left(\frac{m\dot{x}^2}{2}+\frac{k_x x^2}{2}\right)=\mu_0\left(F_{N,0}+\Delta F\cdot\cos\omega t\right)\Delta v\cos\omega t\cdot\mathrm{sign}(v_0-\Delta v\cos\omega t). \tag{31}$$

If $\Delta v < v_0$, then the average value of the right-hand side is positive, and the energy of the system is monotonously increasing from one period to the next. However, if $\Delta v > v_0$ then the amplitude of oscillation stabilizes at the value for which the average change in energy during one period vanishes

$$\left\langle \left( F_{N,0} + \Delta F \cdot \cos \omega t \right) \Delta v \cos \omega t \cdot \text{sign}(v_0 - \Delta v \cos \omega t) \right\rangle = 0 \tag{32}$$

where $\langle ... \rangle$ means averaging over one period of oscillation. During one oscillation period, there is a time interval $\tau_1 < \tau < \tau_2$ where $v_0 - \Delta v \cos \tau < 0$:

$$\tau_{1,2} = \pm \tau^* = \pm \arccos(v_0 / \Delta v) \tag{32}$$

Assuming that the oscillation amplitude $\Delta v$ exceeds the mean sliding velocity $v_0$ only slightly, $\tau^*$ can be approximated by

$$\tau^* \approx \sqrt{2(1 - v_0 / \Delta v)}. \tag{32}$$

In this approximation, the condition (32) can be written, after some simple transformations, as $-4(F_{N,0} + \Delta F)\sqrt{2(1 - v_0 / \Delta v)} + \pi \Delta F = 0$. For the ratio of sliding velocity and oscillation velocity amplitude we finally find

$$\frac{v_0}{\Delta v} = 1 - \frac{1}{2} \left( \frac{\pi \Delta F}{4(F_{N,0} + \Delta F)} \right)^2. \tag{32}$$

Let us now calculate the macroscopic coefficient of friction. It is given by the equation

$$\mu = \mu_0 \frac{\left\langle \left( F_{N,0} + \Delta F \cdot \cos \tau \right) \text{sign}(v_0 - \dot{x}) \right\rangle}{F_{N,0}} = \frac{\mu_0}{2\pi} \left[ -4 \int_0^{\tau^*} \left( 1 + \frac{\Delta F}{F_{N,0}} \cdot \cos \tau \right) d\tau + 2\pi \right], \tag{32}$$

which, assuming sufficiently small $\tau^*$ and considering (32) and (32), leads to the equation

$$\frac{\mu}{\mu_0} \approx 1 - \frac{\Delta F}{2 F_{N,0}}. \tag{32}$$

Comparing this with numerical results (Fig. 7) shows that the obtained approximation describes the plateau value of the COF in the resonant case very well.

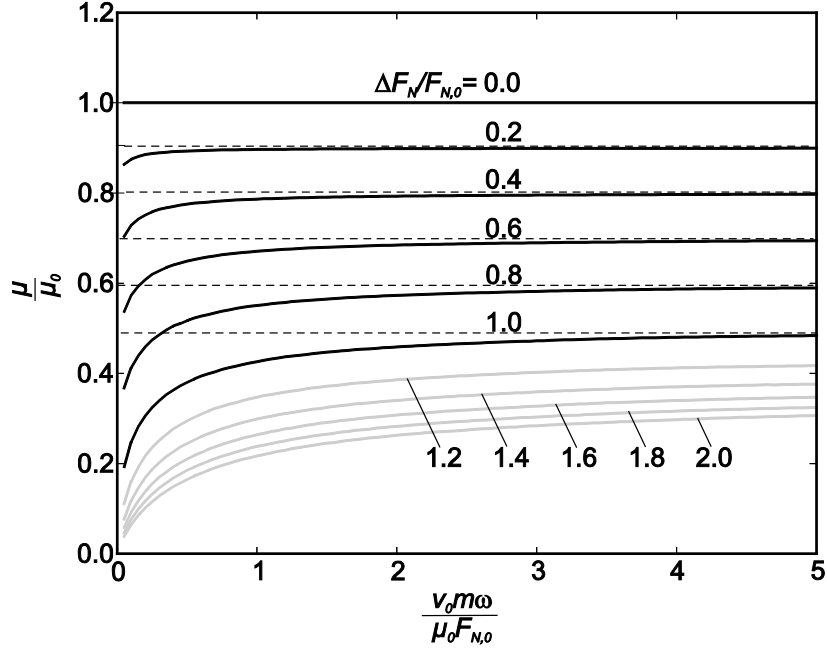

**Fig. 6** The dependence of $\mu_{\text{macro}}/\mu_0$ on the dimensionless velocity $v_0 m\omega/(\mu_0 F_{N,0})$ for the resonant case. The curves start at zero velocity at the static value $\mu/\mu_0 = 1 - \Delta F_N/F_{N,0}$ and tend to the limiting value $\mu/\mu_0 = 1 - \Delta F_N/(2F_{N,0})$ given by Eq. (32) at large velocities. The black curves correspond to the non-jumping case $\Delta F_N/F_{N,0} \leq 1$, and the gray curves to jumping conditions ($\Delta F_N/F_{N,0} > 1$).

## 6 Large oscillation amplitudes ("jumping")

If the amplitude of normal oscillation, $\Delta u_z$, exceeds the average indentation depth $u_{z,0}$, the body starts to "jump": For part of the oscillation period, it will be in contact with the substrate and out of contact the rest of the time. In previous studies this case has not usually been studied in detail. In the first part of this publication [19] the jumping case was also considered (in the context of a stiff system) and it was found that the general character of the dependence of the coefficient of friction on dimensionless sliding velocity is very similar between the jumping and non-jumping cases: In both cases there is a critical velocity above which the COF no longer depends on velocity. Also, the shape of the velocity-dependences changes little after exceeding the critical oscillation amplitude ($\Delta u_z = u_{z,0}$). In analogy to Fig. 3 we present the different dependences for the jumping case in Fig. 7. Comparison with the corresponding graph for critical amplitude $\Delta u_z = u_{z,0}$ presented in Fig. 3b shows that the general character of the dependences remains roughly the same. In particular, in the resonant case IV considered above (corresponding to $\alpha = 1$) there is still a plateau. However, the level of the plateau decreases with increasing the oscillation amplitude.

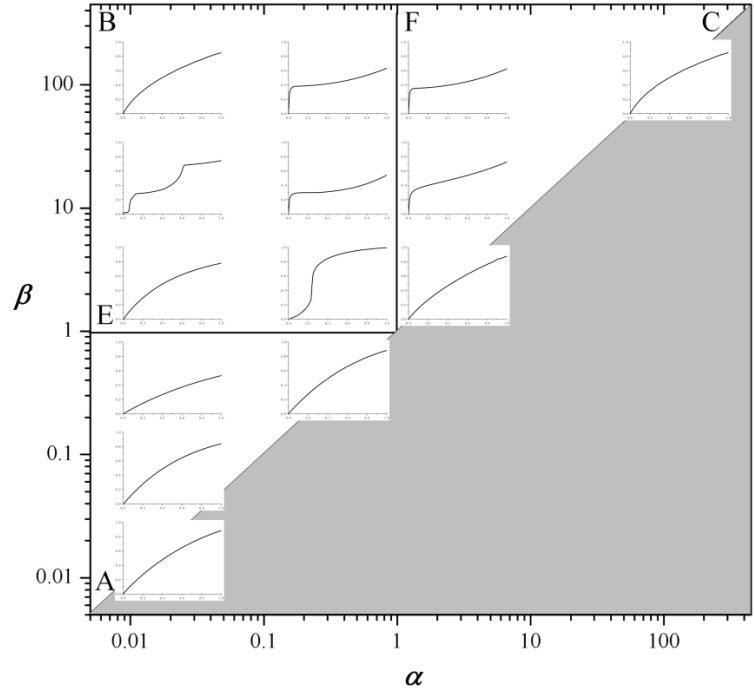

**Fig. 7** Typical dependencies of the reduced coefficient of friction $\mu_{\text{macro}}/\mu_0$ on the dimensionless velocity $\bar{v}$ for the the relative oscillation amplitude $\Delta u_z / u_{z,0} = 1.5$ (jumping case).

## 7 Comparison with experiment in the resonant case

Of the various cases considered in the above discussion, several were studied experimentally in the past. The case I of a stiff system (or high-frequency oscillation) was studied experimentally e.g. in [16]. On the other hand, we are not aware of previous experiments for the resonant case IV. We therefore conducted experiments using a pin-on-disc tribometer (Fig. 8). The natural frequency of pin was determined by impacting the pin and measuring its damped oscillation with a laser vibrometer (Fig. 8b).

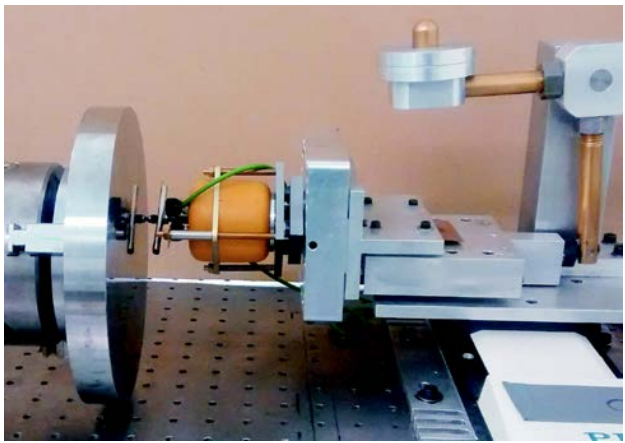
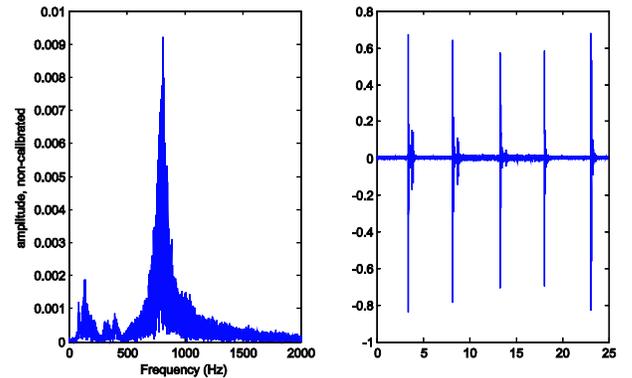

(a)                 (b)

**Fig. 8** (a) a photograph of the experimental set-up: a pin-on-disc tribometer is equiped with an electromagnetic shaker producing out-of-plane oscillations at the resonant frequancy of the pin. (b) The resonant frequency was determined by impacting the pin in tangential direction and determining the Fourier spectrum of the response. The measured natural frequency of the pin was about 800 Hz.

As the determined natural frequency was around 800 Hz, the usual method of exciting oscillations with built-in piezo-elements could not be used, and the tribometer was extended with an electromagnetic shaker as shown in Fig. 8a. The frequency of the shaker was tuned to the natural frequency of the pin, thus creating the conditions of the resonant case IV. The results are presented in Fig. 9. In contrast with non-resonant cases, where the COF increases monotonically with increasing velocity, in the resonant case it was approximately constant (within the relatively large stochastic error).

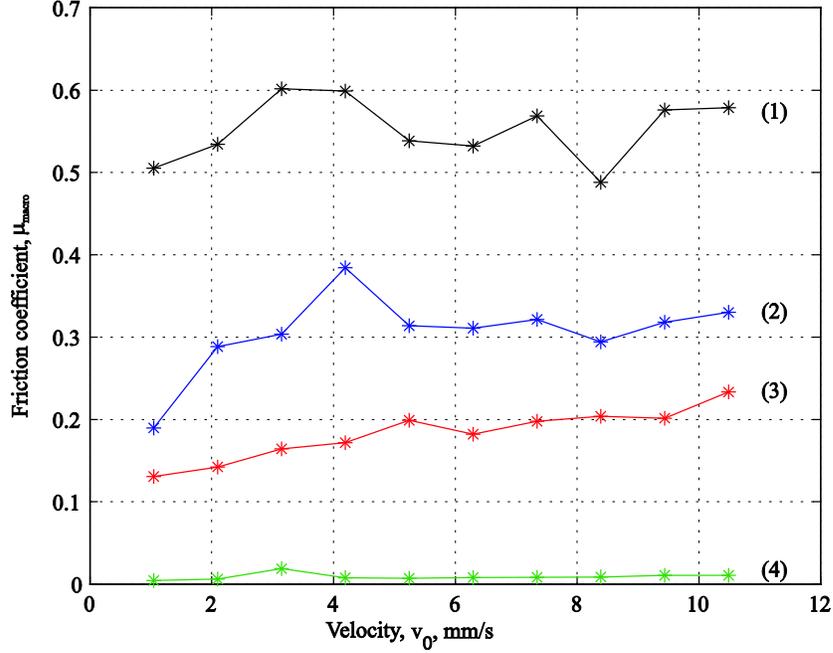

**Fig. 9** Dependence of the coefficient of friction on velocity fo the resonant case. The oscillation amplitudes were: (1) 1.3 μm ; (2) 5.4 μm ; (3) 8.2 $\mu m$ ; (4) 60 μm .

## 8 Summary

We presented a general theoretical analysis of the influence of out-of-plane oscillations on the macroscopically observed coefficient of friction. Unlike previous works, we explicitly took into account both the contact stiffness and the stiffness of the measuring system.

The main governing parameters of the resulting system appear to be the ratios of two natural frequencies of the system (one related to the contact stiffness of the system and the other to combined stiffness of the system and contact) to the frequency of the normal oscillation. As observed in previous works, the velocity-dependence of the COF was found to have two main reference points:

(1) The value at vanishing sliding velocity (static coefficient of friction), which naturally does not depend on the dynamic properties and is solely determined by the smallest normal force during the oscillation cycle.

(2) The characteristic velocity above which the COF no longer depends on the sliding velocity and is equal to its microscopic value $\mu_0$.

The only exceptions from this rule are the two resonant cases. One where the COF is constant and equal to $\mu_0$ at all velocities (III) and a second case where the oscillation frequency is equal to the natural frequency of of the pin. In this latter case the COF tends to a plateau value below $\mu_0$ and does not have a maximum velocity above which the reduction of the COF disappears. To the best of our knowledge, this resonance case was not studied yet and is described here for the first time.

Fig. 10 summarizes schematically the main findings of the present paper. Contrary to the previous figures, we use the non-normalized coefficient of friction and the non-normalized sliding velocity $v_0$, as this better highlights the main tendencies and is easier to compare with experiment.

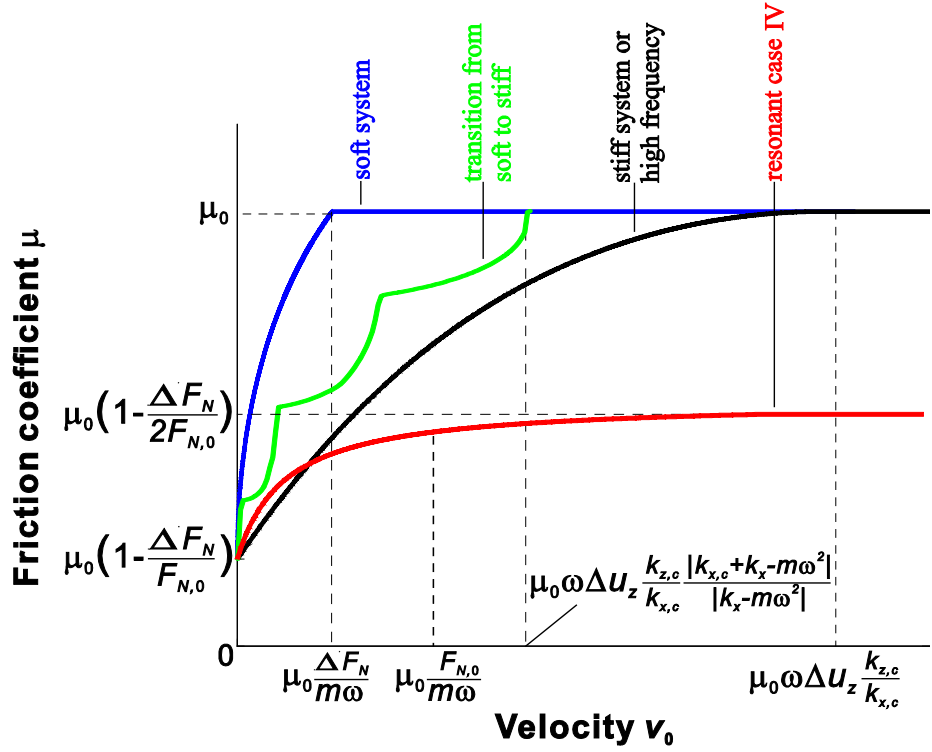

**Fig. 10** Schematic representation of the law of frition (dependence of the friction coefficient on the macroscopic sliding velocity) for different relations between the contact and system stiffness as well as eigenfrequencies and the oscillation frequency.

All dependencies of the macroscopically observed coefficient of friction start at the same static value $\mu_0\left(1-\Delta F_N / F_{N,0}\right)$, which is determined by the smallest normal force during the oscillating cycle. The further shape of the law of friction depends strongly on the dynamical properties of the system.

The case of the very soft system (and stiff contact), which was studied theoretically in [7] and [14], is shown in Fig. 10 with a blue line. In this case, the coefficient of friction first increases very rapidly from the static value, achieves the macroscopic value $\mu_0$ at the critical velocity $\mu_0 \Delta F_N / m\omega$ and does not further change with increasing velocity. The critical velocity, in this case, depends solely on the inertial properties of the system, but not on its stiffness. However, in this approximation the theoretical predictions showed poor fit with experimental data [7]. According to [7] a much better fit to experimental data is achieved if the contact stiffness is taken into account.

The case of finite contact stiffness and very rigid measuring system was considered in detail in the first part of this series [19] and is represented in Fig. 10 with a black curve. The curve starts at the same static value $\mu_0\left(1-\Delta F_N / F_{N,0}\right)$ of the COF and increases with increasing velocity, however not as rapidly as in the case of the soft system. After reaching the value $\mu_0$ at the critical ve-locity $\mu_0 \omega \Delta u_z k_{z,c} / k_{x,c}$, it remains constant. In this case the critical velocity does not depend on inertial properties of the system. However, the contact stiffness also does not enter explicitly into the critical velocity; only the ratio of the normal and tangential stiffness (the Mindlin ratio) appears in the equation. This ratio only depends on the Poisson ratio of the contacting partners and is equal to 1.25 for the typical case of $v = 1/3$. As shown in this paper, this case is also applicable at very high oscillation frequencies independently of contact and system stiffness.

The law of friction in the transition region between soft and stiff system is schematically represented by the green curve in Fig. 10. In the transition region the dependencies of the coefficient of friction on the sliding velocities can have a complicated shape and are sensitive to the parameters of the system and the frequency of oscillations (see Fig. 5 and Fig. 6). Regardless of this complexity, all curves start at the same static friction value $\mu_0 \left(1 - \Delta F_N / F_{N,0}\right)$ and reach $\mu_0$ at the critical velocity given by Eq. (12). Depending on parameters, this velocity can range from zero to infinity.

When approaching the resonant case IV where the frequency of the external oscillation is equal to the natural frequency of the system, the critical velocity tends to infinity and the COF reaches a plateau value less than $\mu_0$. For the exactly resonant case, the COF does not exceed the value $\mu_0 \left(1 - \Delta F_N / (2F_{N,0})\right)$, which is larger than the static value $\mu_0 \left(1 - \Delta F_N / F_{N,0}\right)$ but smaller than $\mu_0$ even at very high sliding velocities.

In conclusion, we would like to stress once again that the entire analysis of this paper is based on the assumption that Coulomb's law of friction with a constant coefficient of friction is valid locally, in the immediate contact point. We have shown that the macroscopic behavior can be very non-trivial despite the simplicity of the underlying local law of friction. However, a more general analysis taking into account system dynamics, contact stiffness and changes of local friction may eventually achieve the best fit with experimental data. However, we believe that changes in the local COF will not impact the overall classification of the discussed dynamic cases.

## 9 Acknowledgements

The authors would like to thank J. Wallendorf and Qiang Li for their help with preparing figures for the paper. This work was supported in part by Tomsk State University Academic D.I. Mendeleev Fund Program, grant No. 8.2.19.2015.The authors would like to thank J. Wallendorf and Qiang Li for their help with preparing figures for the paper. This work was supported in part by Tomsk State University Academic D.I. Mendeleev Fund Program, grant No. 8.2.19.2015.

## 10 References

[1] Siegert K, Ulmer J. Superimposing Ultrasonic Waves on Tube and Wire Drawing. *J Eng Mater Technol* **123**(724): 517–523 (2000)

[2] Egashira K, Mizutani K, Nagao T. Ultrasonic vibration drilling of microholes in glass. *CIRP Ann Technol* **51**(1): 339–342 (2002)

[3] Godfrey D. Vibration reduces metal to metal contact and causes an apparent reduction in friction. *ASLE Trans* **10**(2): 183–192 (1967)

[4] Fridman H D, Levesque P. Reduction of static friction by sonic vibrations. *J Appl Phys* **30**(10): 1572–1575 (1959)

[5] Broniec Z, Lenkiewicz W. Static friction processes under dynamic loads and vibration. *Wear* **80**(3): 261–271 (1982)

[6] Littmann W, Storck H, Wallaschek J. Sliding friction in the presence of ultrasonic oscillations: superposition of longitudinal oscillations. *Arch Appl Mech* **71**(8): 549–554 (2001)

[7] Teidelt E. *Oscillating contacts: friction induced motion and control of friction.* Ph.D. Thesis. Berlin (Germany): Technische Universität Berlin, 2015.

[8] Popov V L, Wetter R. Symmetry breaking as a general design principle of oscillation-based methods for fixation and manipulation of nano-objects. *Advanced Biomaterials and Devices in Medicine*, **3**(1): 10-18 (2016)

[9] Weishaupt W. Reibungsverminderung durch mechanische Schwingungen. *Tech Mess* **11**: 345–348 (1976)

[10] Goto H, Ashida M, Terauchi Y. Effects of ultrasonic vibration on the wear characteristics of a carbon steel: analysis of the wear mechanism. *Wear* **94**(1): 13–27 (1984)

[11] Chowdhury M A, Helali M M. The effect of frequency of vibration and humidity on the coefficient of friction. *Tribol Int* **39**(9): 958–962 (2006)

[12] Chowdhury M A, Helali M M. The effect of amplitude of vibration on the coefficient of friction for different materials. *Tribol Int* **41**(4): 307–314 (2008)

[13] Popov V L, Starcevic J, Filippov A E. Influence of ultrasonic in-plane oscillations on static and sliding friction and intrinsic length scale of dry friction processes. *Tribol Lett*: **39**(1): 25-30 (2010)

[14] Teidelt E, Starcevic J, Popov V L. Influence of ultrasonic oscillation on static and sliding friction. *Tribol Lett* **48**(1): 51–62 (2012)

[15] Milahin N, Li Q, Starcevic J. Influence of the normal force on the sliding friction under ultrasonic oscillations. *Facta Univ Ser Mech Eng* **13**(1): 27–32 (2015)

[16] Milahin N. *Robuste Einflussparameter für Reibung und Oberflä-chenmodifizierung unter Einfluss von Ultraschall*. Ph.D. Thesis. Berlin (Germany): Technische Universität Berlin, 2016.

[17] Nguyen H X, Teidelt E, Popov V L, Fatikow S. Dynamical tangential contact of rough surfaces in stick-slip microdrives: modeling and validation using the method of dimensionality reduction. *Physical Mesomechanics* **17**(4): 304-310 (2014)

[18] Cabboi A, Putelat T, Woodhouse J. The frequency response of dynamic friction: enhanced rate-and-state models. *Journal of the Mechanics and Physics of Solids* **92**: 210–236 (2016)

[19] Popov M, Popov V L, Popov N V. Reduction of friction by normal oscillations. I. Influence of contact stiffness. *Friction,* 2016 submitted.